# Exciton-Plasmon Coupling Effects on the Nonlinear Optical Susceptibility of Hybrid Quantum Dot-Metallic Nanoparticle System

Nam-Chol Kim,[†,*] Myong-Chol Ko,[†] Song-Jin Im,[†] Zhong-Hua Hao[‡]

[†] Department of Physics, Kim Il Sung University, Pyongyang, D. P. R. Korea

[‡] Department of Physics, Key Laboratory of Acoustic and Photonic Materials and Devices of Ministry of Education, Wuhan University, Wuhan 430072, P. R. China

**Abstract:** We have studied theoretically the exciton-plasmon coupling effects on the third-order optical nonlinearity of a coherently coupled hybrid system of a metal nanoparticle (MNP) and a semiconductor quantum dot (SQD) in the presence of a strong control field with a weak probe field. We deduced the analytic formulas of exciton population dynamics and the effective third-order optical susceptibility and numerically analyzed the nonlinear optical response of the hybrid system which is greatly enhanced due to exciton–plasmon couplings. Our results show that one can manipulate the nonlinear optical absorption and refraction by controlling the frequency of the control light or varying the intensity and the direction of the light and adjusting the interparticle distance. The results obtained here may have the potential applications of nanoscale optical devices such as optical switches.

E-mail: elib.rns@hotmail.com

# 1. Introduction

The hybrid nanostructures based on metal nanoparticles (MNPs) and semiconductor quantum dots (SQDs) are the area of considerable current interest[1-3]. Especially the interaction between exciton systems and plasmonic structures have attracted considerable attention. The hybrid supernanostructures, MNP-SQD hybrid system show many interesting phenomena, such as plasmon-induced fluorescence enhancement and quenching[4], plasmon-assisted Forster energy transfer[5], induced exciton-plasmon-photon conversion [6], generation of a single plasmons [7], modifying the spontaneous emission in semiconductor quantum dots (SQDs) [8], etc. The hybrid systems of plasmonics coupled with semiconductor dye molecules, or quantum dots have been intensively studied due to their extraordinary properties such as the enhancement of radiative emission rates, absorption of the exciton light and non-radiative energy transfer [9–11]. Many investigations of the third-order optical nonlinearity that arise in the coherently coupled exciton-plasmon system have also been reported. Recently, Wang and his co-workers experimentally demonstrated that the effective nonlinear absorption and refraction of CdS-Ag core-shell QDs comparing with CdS QDs were greatly enhanced [12]. Furthermore, Ji's group observed the intensity-dependent enhancement of saturable absorption in a PbS-Au nanohybrid composite [13]. In theoretical research, Xiong *et al*. reported that third-order nonlinear optical susceptibility of CdTe QDs in the hybrid SQD-MNP system can be effectively modified by local field enhancement and dipole interaction [14]. Enhancement of Kerr nonlinearity in the MNP-SQD hybrid system has also been theoretically demonstrated [15]. Zhu and his co-workers have theoretically demonstrated the ways of plasmon-assisted mass sensing based on the hybrid SQD-MNP system[16]. Quantum coherence and interference phenomena in a quantum dot (SQD)-metallic nanorod (MNR) hybrid system have also been studied in Ref [17] and the Purcell effect, acceleration of a spontaneous emission recombination rate, has been observed in InN/In nanocomposites with buried nanoparticles of metallic In [18]. Despite such studies, SQD–MNP hybrid systems seem to be the host of many more interesting phenomena to explore. Because of several extraordinary properties of the transmission of light, such as high light speed, low loss and no interference during transmission, photonics has overwhelming benefits over electronics to process information and to perform switching effects, and has been pursued for the advanced technology of the future. Nonlinear optics plays an important role in photonics and quantum electronic devices. The main difficulty in achieving these applications is that conventional materials offer only a very small nonlinear optical response, which is significantly outweighed by linear absorption. In order to seek the ideal nonlinear optical materials, there is a great deal of work devoted to identifying the nonlinear optical response in a wide variety of materials [19–23]. But most of the nonlinear optical materials mentioned above only focus on the conventional semiconductor materials. The investigation of optical nonlinearities in a hybrid complex is still lacking. The hybrid system of nanoparticles is obviously different from conventional low-dimensional semiconductors [24–29]. Colloidal nanoparticles confine carriers strongly and do not permit efficient tunnel coupling. The interparticle Coulomb interaction between the nanoscale building blocks becomes the main coupling mechanism and can substantially change the physical properties of these nanoscale building blocks [24]. Consequently, it is very important to investigate the optical nonlinearity in a hybrid complex system.

In this Letter, we will study the exciton-plasmon coupling effects on the third-order nonlinear optical susceptibility in the hybrid MNP-SQD system theoretically based on the method of Ref [25]. The basic excitations in the MNP are the surface plasmons with a continuous spectrum. In SQDs, the excitations are the discrete interband excitons. When the exciton energy in a SQD lies in the vicinity of the plasmon peak of the MNP, the coupling of the plasmon and exciton becomes very strong [9]. The exciton coherent dynamics are modified strongly by the strong coherent interaction.

## 2. Model and Theory

We now consider a hybrid molecule composed of a spherical MNP of radius $R$ and a spherical SQD with radius $r$ in the presence of polarized external field $E = E_0 \cos(\omega t)$, where the direction of polarization is specified below. Because of its symmetry, a spherical SQD has three bright excitons with optical dipoles parallel to the direction α, where α can be x, y, and z[9]. The center-to-center distance between the two nanoparticles is $d$ (Fig. 1(a)). The hybrid is subject to a strong control field and a weak signal field (Fig. 1(b)). The SQD with exciton ground state |0> and exciton state |1> is described by a density matrix and the MNP by classical electrodynamics.

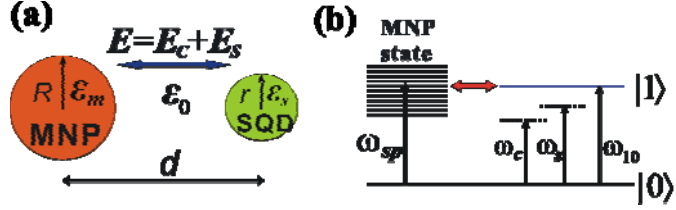

Fig. 1 (Color online). (a) Schematic diagram of the hybrid system driven by a strong control field with amplitude $E_c$ and frequency $\omega_c$ and probed by a weak signal field with amplitude $E_s$ and frequency $\omega_s$ ($E_c \gg E_s$). $r$ and $R$ are the radii of SQD and MNP, respectively. $d$ is the center-to-center distance between SQD and MNP. $\varepsilon_0$, $\varepsilon_s$ and $\varepsilon_m$ are the dielectric constants of the background medium, SQD and MNP, respectively. (b) Energy level diagram of the system. $\omega_{10}$ and $\omega_{sp}$ are the frequencies of exciton and surface plasmon, respectively.

The Hamiltonian of the system in the rotating-wave approximation can be written as

$$H = \hbar\Delta\hat{\sigma}_z - \mu\left(\tilde{E}_{SQD}\hat{\sigma}_{10} + \tilde{E}^*_{SQD}\hat{\sigma}_{01}\right) \quad (1)$$

where $\hat{\sigma}_{ij}$ ($i, j = 0, 1$) is the dipole transition operator between $|i>$ and $|j>$ of SQD and $\hat{\sigma}_z = (\hat{\sigma}_{11} - \hat{\sigma}_{00})/2$. $\Delta = \omega_{10} - \omega_c$ is the difference between the exciton frequency $\omega_{10}$ and the control field frequency $\omega_c$. $\tilde{E}_{SQD}$ and $\tilde{E}_{MNP}$ are the total electric field felt by the SQD and MNP, respectively. Here, $\tilde{E}_{SQD} = E_c + E_s e^{-i\delta t} + S_\alpha P_{MNP} / [3\varepsilon_{eff} d]$, $\tilde{E}_{MNP} = E_c + E_s e^{-i\delta t} + S_\alpha P_{SQD} / [3\varepsilon_{eff} d]$ [9, 26]. $\varepsilon_{eff} = (2\varepsilon_0 + \varepsilon_s)/3$, where $\varepsilon_0$ and $\varepsilon_s$ are the dielectric constants of the background medium and SQD, respectively. $\delta = \omega_s - \omega_c$ is the frequency difference of the signal field and the control field. Geometric factor $S_\alpha$ is equal to 2 (−1) when $E$ is parallel to the $z$ ($x, y$) axis. The z direction corresponds to the axis of the hybrid system. The dipole of the MNP $P_{MNP} = \alpha_{MNP}\tilde{E}_{MNP}$ derives from the charge induced on the surface of the MNP, where $\alpha_{MNP} = \varepsilon_0 R^3 (\varepsilon_m - \varepsilon_0)/(2\varepsilon_0 + \varepsilon_m)$ [27]. The dipole moment of the SQD is $P_{SQD} = \mu\sigma_{10}$ [25]. By ignoring the quantum properties of $\hat{\sigma}_{10}$ and $\hat{\sigma}_z$ [28], the temporal evolutions of the exciton in the SQD are determined by the Heisenberg equation of motion as

$$\frac{d\hat{\sigma}_{01}}{dt} = -i\Delta\hat{\sigma}_{01} - 2i\frac{\mu(E_{SQD} + E_s e^{-i\delta t})}{\hbar}\hat{\sigma}_z \quad (2)$$

$$\frac{d\hat{\sigma}_z}{dt} = i\frac{\mu}{\hbar}\left(E_{SQD}\hat{\sigma}_{10} - E^*_{SQD}\hat{\sigma}_{01}\right) + i\frac{\mu}{\hbar}\left(E_s e^{-i\delta t}\hat{\sigma}_{10} - E^*_{SQD}e^{i\delta t}\hat{\sigma}_{01}\right) \quad (3)$$

Based on the Heisenberg equation of motion, if we set $p = \mu\sigma_{10}$ and $w = 2\sigma_z$, we have the following equations as

$$\frac{dp}{dt} = \left(-\frac{1}{T_2} - i\Delta\right)p - \frac{i\mu^2}{\hbar}wE, \quad (4)$$

$$\frac{dw}{dt} = -\frac{1}{T_1}(w+1) + \frac{4}{\hbar}\text{Im}(pE^*). \quad (5)$$

In order to solve the Eqs. (4) and (5), we make the ansatz [25]

$$p = p_0 + p_1 e^{-i\delta t} + p_{-1}e^{i\delta t}, \quad (6)$$

$$w = w_0 + w_1 e^{-i\delta t} + w_{-1}e^{i\delta t}. \quad (7)$$

Substituting Eqs. (6) and (7) into Eqs. (4) and (5) results in a set of steady state equations as

$$0 = -\left(\frac{1}{T_2} + i\Delta\right)p_0 - \frac{i\mu^2 B}{\hbar}w_0 p_0 - \frac{i\mu^2 A}{\hbar}w_0 E_c, \quad (8)$$

$$-i\delta p_1 = -\left(\frac{1}{T_2} + i\Delta\right)p_1 - \frac{i\mu^2 B}{\hbar}(w_0 p_1 + w_1 p_0) - \frac{i\mu^2}{\hbar}(Aw_1 E_c + w_0 E_s), \quad (9)$$

$$i\delta p_{-1} = -\left(\frac{1}{T_2} + i\Delta\right)p_{-1} - \frac{i\mu^2 B}{\hbar}(w_0 p_{-1} + w_{-1} p_0) - \frac{i\mu^2 A}{\hbar}w_{-1}E_c, \quad (10)$$

$$0 = -\frac{1}{T_1}(w_0 + 1)p_0 + \frac{2i}{\hbar}A(p_0^* E_c - p_0 E_c), \quad (11)$$

$$-i\delta w_1 = -\frac{1}{T_1}w_1 + \frac{2i}{\hbar}\left[A(p_{-1}^* E_c - p_1 E_c^*) + p_0^* E_s\right], \quad (12)$$

$$i\delta w_{-1} = -\frac{1}{T_1}w_{-1} + \frac{2i}{\hbar}\left[A(p_1^* E_c - p_{-1}E_c^*) - p_0 E_s^*\right], \quad (13)$$

where $A = 1 + \frac{\gamma R^3 S_\alpha}{\varepsilon_{eff1}d^3}$, $B = \frac{\gamma R^3 S_\alpha^2}{\varepsilon_{eff1}\varepsilon_{eff2}d^6}$. By solving the Eqs. (8)-(13), we obtained the following Eqs as

$$w_{-1} = -\frac{A\mu^2 w_0 E_s^* E_c T_2^2}{\hbar^2 D(\delta_c)}(2 + i\delta_c)[1 - i(\Delta_c + B_c w_0)][1 + i(\delta_c + \Delta_c + B_c w_0)], \quad (14)$$

$$p_{-1} = \frac{iA^2\mu^4 w_0 E_s^* E_c^2 T_2^3}{\hbar^3 D(\delta_c)[1+i(\Delta_c + B_c w_0)]}(1+i\Delta_c)(2+i\delta_c)[1-i(\Delta_c + B_c w_0)], \quad (15)$$

where $N$ is the number density of the hybrid system, $\Omega_c^2 = \mu^2|E_c|^2 T_2^2/\hbar^2$ is the generalized intensity of the control field, $\delta_c = \delta T_2$, $\Delta_c = \Delta T_2$ and $B_c = \mu^2 B T_2/\hbar^2$. Therefore the nonlinear optical susceptibility can read as follows:

$$\chi_{eff}^{(3)} = \frac{N p_{-1}}{3 E_c^2 E_s^*} = \frac{i N A^2 \mu^4 w_0 T_2^3}{3\hbar^3 D(\delta_c)[1+i(\Delta_c + B_c w_0)]}(1+i\Delta_c)(2+i\delta_c)[1-i(\Delta_c + B_c w_0)] \quad (16)$$

where

$$D(\delta_c) = 2\left(i\delta_c + \frac{T_2}{T_1}\right)\left[1+(\Delta_c + B_c w_0)^2\right]\left[(1+i\delta_c)^2 + (\Delta_c + B_c w_0)^2\right]$$

$$+ 2A^2\Omega_c^2\left[(1+\Delta_c^2)(1+i\delta_c) - B_c w_0(B_c w_0 - i\delta_c\Delta_c)\right], \quad (17)$$

Im$\chi_{eff}^{(3)}$ and Re$\chi_{eff}^{(3)}$ represent the nonlinear absorption and refraction, respectively. The population inversion of the exciton $w_0$ is determined by the cubic equation

$$w_0 = -A^2\Omega(T_1/T_2)w_0/\left[1+(\Delta_c + B_c w_0)^2\right] - 1/4 \quad (18)$$

## 3. Results and Discussion

In the following, we mainly provide some typical parameters for calculation of these complicated expressions. We consider a Au MNP with radius $R = 7.5$nm and the dielectric constants of the background medium and SQD: $\varepsilon_0 = 1$, and $\varepsilon_s = 6$ [15, 29]. For the relaxtion time and dipole moment in SQD, we take $T_1 = 0.8$ ns, $T_2 = 0.3$ ns and $\mu = 10^{-28}$ C·m [9], and $N = 10^{20}$ m$^{-3}$. We also choose the exciton resonant frequency to be 2.5 eV which is near the broad plasmon frequency of gold (peak near 2.4eV with a width of approximately 0.25eV) [9].

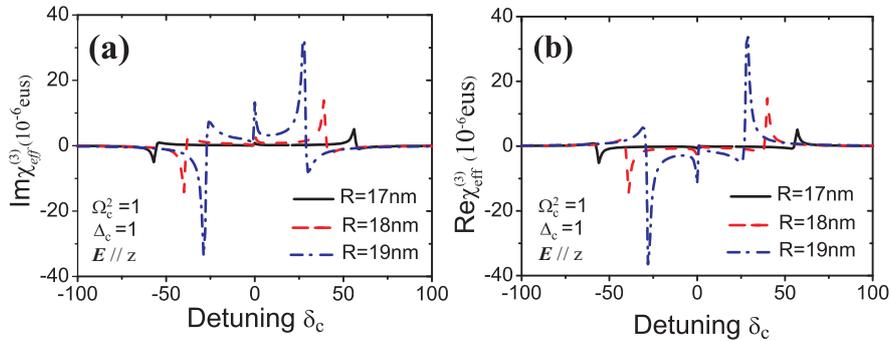

Fig. 2 (Color online). Nonlinear absorption Im$\chi_{eff}^{(3)}$ (a) and refraction Re$\chi_{eff}^{(3)}$ (b) as a function of the detuning $\delta_c = (\omega_s - \omega_c)T_2$ with $\Delta_c = (\omega_{10} - \omega_c)T_2 = 1$ and $R = 17, 18, 19$ nm.

Figures 2(a) and 2(b) show the nonlinear absorption and refraction spectra as a function of the detuning $\delta_c = (\omega_s - \omega_c)T_2$. In Fig. 2(a), there are three peaks of the absorption spectra, each of which is originated from the three-photon resonance (the first peak from the left), stimulated Rayleigh resonance (the second peak from the left) and ac-Stark resonance (the third peak from the left), respectively [25]. In the region of the first peak, the absorption spectrum attributed to the three-photon effect evolves from a Fano-like lineshape into a positive peak. In the region of the second peak, the absorption spectrum assigned to the Rayleigh resonance evolves from a Fano-like lineshape into a positive peak. In region of the third peak, the absorption spectrum induced by the ac-Stark effect evolves from the Fano-like lineshape into a negative peak. The corresponding refraction spectra are shown in Fig. 2(b). From resonant excitation to near-resonant excitation, the dynamic evolution of the absorption and refraction spectra implies that the dominant role of three resonance mechanisms changes continuously. From the results in Figs. 2(a) and 2(b), we can see that nonlinear optical responds are very sensitive to the interparticle distance in the region of short distance.

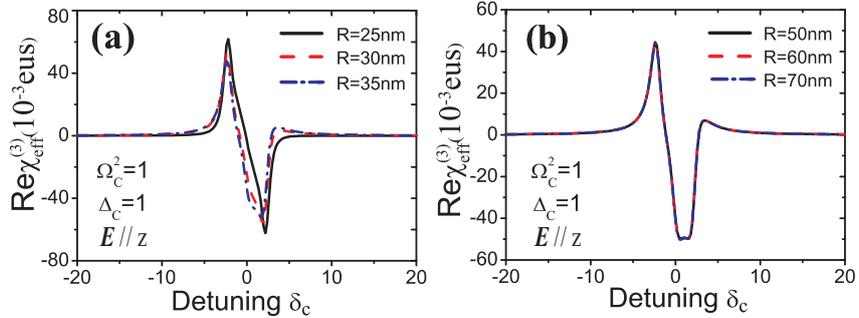

Fig. 3 (Color online). Refraction Re$\chi_{eff}^{(3)}$ as a function of the detuning $\delta_c = (\omega_s - \omega_c)T_2$ with $\Delta_c = 1$ and $\Omega_c^2 = 1$. (a) $R = 25, 30, 35$ nm. (b) $R = 50, 60, 65$ nm.

Figures 3(a) and 3(b) show the refraction spectra as a function of the detuning $\delta_c = (\omega_s - \omega_c)T_2$ in the large interparticle distances. From the Fig. 3(a), we see that the refraction spectra are enhanced in the sample with shorter interparticle distance. The local field resulting from the Coulomb interaction between MNP and SQD counteracts the external field and makes the total field in the media weaker. The shorter the distance between MNP and SQD is, the stronger the Coulomb interaction becomes, thus, resulting in the total field will be weaker. Figure 3(b) show that the refraction nearly doesn't depend on the interparticle distance larger than about 50nm. From the Figs. 2 and 3, we can see that the variation of nonlinear optical response to the interparticle distance vanishes as the interparticle distance exceeds about 50nm.

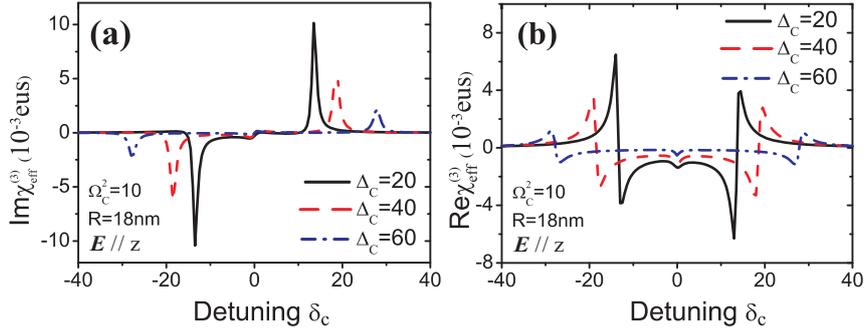

Fig. 4 (Color online). Nonlinear absorption Im$\chi_{eff}^{(3)}$ (a) and refraction Re$\chi_{eff}^{(3)}$ (b) as a function of the detuning $\delta_c = (\omega_s - \omega_c)T_2$ with $R$ = 18nm and $\Delta_c = (\omega_{10} - \omega_c)T_2$ = 20, 40, 60.

Figures 4(a) and 4(b) show the dependence of nonlinear absorption and refraction spectra on the detunings $\Delta_c$. The sample with interparticle distance $R$ = 18 nm is selected and the Rabi frequency of the control light is fixed at $\Omega_c^2$ = 10. Figure 4(a) show the absorption spectra with the detunings $\Delta_c$ =20, 40, 60. From Fig. 4(a), we see that the height of absorption peak is lowered as the detuning is increased. The nonlinearity here is due to the saturation of absorption at high laser powers in a two-level system of semiconductor quantum dot. The role of the metal nanoparticle is to enhance this nonlinearity in the regime of exciton–plasmon resonance. The figures 4(a) and 4(b) show that the maximum and minimum of the nonlinear responses can occur at some values of $\delta_s$ due to the dynamic shift caused by exciton–plasmon coupling.

We can also consider the dependency of nonlinear absorption and refraction with the Rabi frequency or the generalized intensity of the control field $\Omega_c^2$. Figures 5(a) and 5(b) illustrate the nonlinear absorption Im$\chi_{eff}^{(3)}$ and refraction Re$\chi_{eff}^{(3)}$, respectively, as a function of $\delta_s$ for three different Rabi frequencies of the control light when $E$ is parallel to the axis of the hybrid system. Here we choose a sample with interparticle distance $R$ =

18nm and set the detuning frequency of the control light at $\Delta_c = 1$. As we can see easily from the Figs. 5(a) and 5(b), a control light with a weaker Rabi frequency can enhance the nonlinear optical absorption and refraction significantly. We also found that the larger the Rabi frequency is, the further the position of maximum of the nonlinear optical absorption and refraction is placed from the point $\delta_c=0$. When the field $E$ is perpendicular to the axis of the hybrid system, the absorption and refraction are shown in Figs. 5(c) and 5(d), respectively. Figs. 5(c) and 5(d) show that the absorption and refraction are enhanced and the position of maximum of them are closer to the point $\delta_c=0$ than those when $E$ is parallel to the axis of the hybrid system. As required by the nonlinear Kramers-Kronig relation, the complementary behavior between the absorptive and refractive features is evident from Fig. 5.

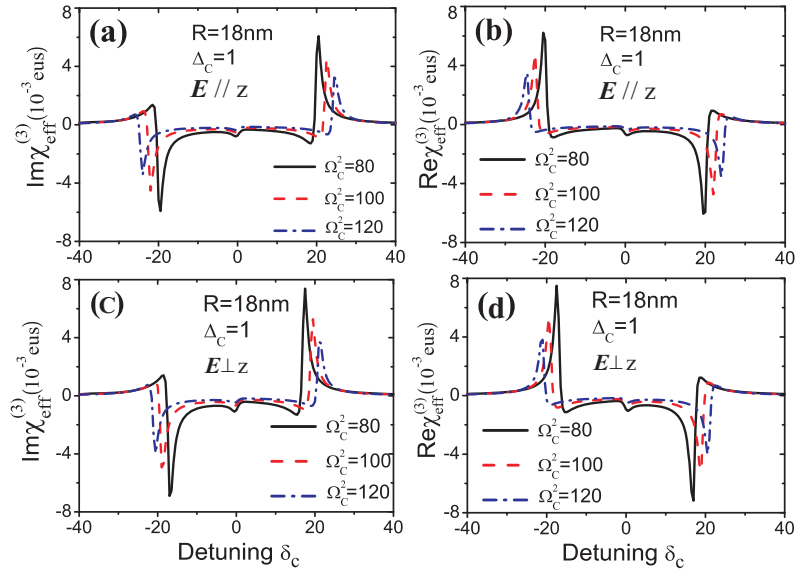

Fig. 5 (Color online). Nonlinear absorption $\text{Im}\chi_{\text{eff}}^{(3)}$ and refraction $\text{Re}\chi_{\text{eff}}^{(3)}$ as a function of the detuning $\delta_c = (\omega_s - \omega_c)T_2$. (a) Absorption $\text{Im}\chi_{\text{eff}}^{(3)}$ and (b) refraction $\text{Re}\chi_{\text{eff}}^{(3)}$ when $E$ is parallel to the axis of the hybrid system. (c) Absorption $\text{Im}\chi_{\text{eff}}^{(3)}$ and (d) refraction $\text{Re}\chi_{\text{eff}}^{(3)}$ when $E$ is perpendicular to the axis of the hybrid system. We set $R = 18\text{nm}$, $\Delta_c = (\omega_{10} - \omega_c)T_2 = 1$ and $\Omega_c^2 = 80, 100, 120$.

Finally, we consider the polarization dependence of nonlinear absorption of the MNP - SQD hybrid system (see Fig. 6). From figure 6(a), we see that the absorption is enhanced when $E \perp z$ than $E \parallel z$, where we set $R = 18$ nm, $\Delta_c = 1$, $\Omega_c^2 = 100$. On the contrary, the absorption can be suppressed strongly when $E \perp z$ than $E // z$ by setting the parameters as $R = 18$ nm, $\Delta_c = 20$, $\Omega_c^2 = 10$. From this result, we can say that by changing the parameters such as the detuning of control light or probe light from the exciton frequency, the interparticle distance, Rabi frequency and the orientation of the electric field applied to the system, one can obtain the different nonlinear absorption and refraction. The physics cause of this enhanced or suppressed response is the local field resulting from the Couloumb interaction between MNP and SQD. The effective field applied to MNP and SQD is the superposition of the external field and the induced internal field. Enhancement or suppression of the effective field depends on the polarization. Therefore, the polarization dependence shown in Figs. 6(a) and 6(b) is also a result of interference of the external field and the induced internal field.

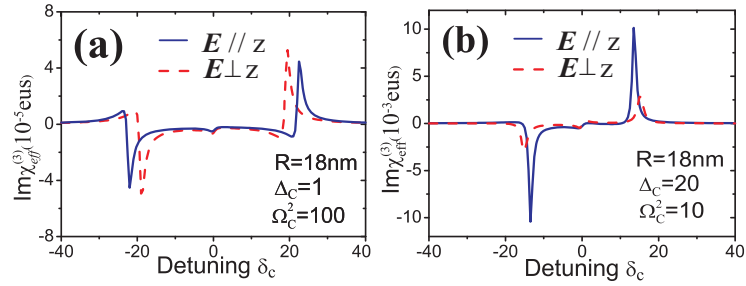

Fig. 6 (Color online). Comparison of nonlinear absorptions $\mathrm{Im}\chi_{\mathrm{eff}}^{(3)}$ as a function of the detuning $\delta_c$ when the electric field $E$ is parallel to the axis of the hybrid system( blue solid line) and when the electric field $E$ is perpendicular to the axis of the hybrid system( red dashed line). (a) $R = 18$ nm, $\Delta_c = (\omega_{10} - \omega_c)T_2 = 1$, $\Omega_c^2 = 100$. (b) $R = 18$ nm, $\Delta_c = (\omega_{10} - \omega_c)T_2 = 20$, $\Omega_c^2 = 10$.

## 4. Conclusion

In conclusion, we have studied theoretically the third-order optical nonlinearity of a coherently coupled MNP-SQD system in the presence of a strong control field with a weak probe field. We deduced the analytic formulas of exciton population and the effective third-order optical susceptibility and numerically analyzed the nonlinear optical response of the hybrid system. We showed that one can manipulate the nonlinear optical absorption and refraction by controlling the frequency of the control light from the exciton frequency or varying the intensity and the direction of the light and adjusting the interparticle distance. The results obtained here may have the potential applications of nanoscale optical devices such as optical switches. We hope that the results presented in this work can be experimentally demonstrated in the near future.

**Acknowledgments.** This work was supported by Key Project for Frontier Research on Quantum Information and Quantum Optics of Ministry of Education of D. P. R of Korea.